\documentclass[11pt,a4paper]{article}
\usepackage[margin=0.9in]{geometry}
\usepackage[utf8]{inputenc}
\usepackage[T1]{fontenc}
\usepackage{authblk}
\usepackage{amsmath,amsthm,amssymb,amsfonts}
\usepackage{mathtools}
\usepackage{comment}
\usepackage{fancyhdr}
\usepackage{sectsty}
\usepackage[english]{babel}
\usepackage{multicol,caption}
\usepackage{url}
\usepackage{graphicx}
\usepackage{wrapfig}
\usepackage{secdot}
\usepackage{gensymb}

\sectionfont{\fontsize{10}{12}\selectfont{}{}}

\begin{document}

\title{\Large \textbf{ Inkjet Printed Wire-Grid Polarizers for the THz Frequency Range}\vspace{-1ex}}

\author[1]{\normalsize A. Farid}
\author[1]{N. J. Laurita}
\author[2]{B. Tehrani}
\author[2]{J. G. Hester}
\author[2]{M. M. Tentzeris}
\author[1]{N. P. Armitage}

\affil[1]{\small \textit{Department of Physics and Astronomy, The Johns Hopkins University, Baltimore, Maryland 21218, USA}}
\affil[2]{\textit{School of Electrical and Computer Engineering, Georgia Institute of Technology, Atlanta, Georgia 30332-250, USA}\vspace{-7ex}}
\date{}

\maketitle

\begin{abstract}\vspace{-1ex}
We have investigated the use of inkjet printing technology for the production of THz range wire-grid polarizers using time-domain terahertz spectroscopy (TDTS). Such technology affords an inexpensive and reproducible way of quickly manufacturing THz range metamaterial structures. As a proof-of-concept demonstration, numerous thin silver-nanoparticle ink lines were printed using a Dimatix DMP-2831 printer. We investigated the optimal printing geometry of the polarizers by examining a number of samples with printed wires of varying thickness and spacing. We also investigated the polarization properties of multiply-stacked polarizers.
\end{abstract}

\section{INTRODUCTION\vspace{-2ex}}
\normalsize Recent advancements in materials studies and the field of terahertz spectroscopy have led to the widespread use of optics, emitters, and equipment for the THz frequency range. This has resulted in improvements in device performance and has created a demand for the development of less expensive, but superior equipment \cite{thzimportance,terahertzreview}. When probing the properties of a material, polarized electromagnetic radiation allows for the easy isolation of certain information in a transmission or reflection measurement \cite{polarizermathandapp}. Polarizers are an important optical component because the accuracy of measurements can depend on the degree to which the electromagnetic waves are polarized. \par

Due to the longer wavelength of THz spectroscopy, polarizers for the THz range can be made from a wire grid. The most common commercial wire grid polarizers are made from tungsten wire \cite{useofwiregridpolarizers} and they tend to be expensive \cite{thorlabs}.  If a high degree of polarization is required, it is likely that multiple polarizers will need to be used in precise measurements.  Although there have been recent advances in the technology \cite{firstterahertzdevice, micrometer, micrometerr, Nanotubes, analysis}, there is interest in developing cheaper and more effective polarizers for THz spectroscopies. \par

In this paper, we describe a simple and inexpensive printing technique to create polarizers made from lines of silver ink printed on Kapton film. This technique has also been used to print small circuits, electronics \cite{inkjet, printingtech} and - in a related fashion - THz range metamaterial structures \cite{Metamaterials}.  Using time domain terahertz spectroscopy (TDTS), we studied the effect of three parameters on the capability of polarizers: the spacing between the silver ink lines (G), the width of the ink lines (W), and the number of passes with the ink-jet printer (L). We also studied the effect that stacking multiple polarizers had on the degree of polarization (DoP). \par

\section{METHODS AND MATERIALS\vspace{-2ex}}
Numerous proof-of-concept polarizer prototypes were inkjet printed on polyimide substrate (Dupont Kapton 500 HN, 5 mil). Each sample was 1 cm $\times$ 1 cm in size and up to 10 were printed at once. Sun Chemical SunTronic EMD5730 ink was printed with a drop spacing of 20 $\mu m$, and each pattern layer took approximately 2 to 4 minutes to print. After printing, the samples were heated on hot plate at 90$\degree$C for 30 minutes to evaporate the solvent. Once dry, samples were sintered at 200$\degree$C for 2 hours.\par

\begin{figure}[!ht] 
	\center
	\includegraphics[width = 0.67\linewidth]{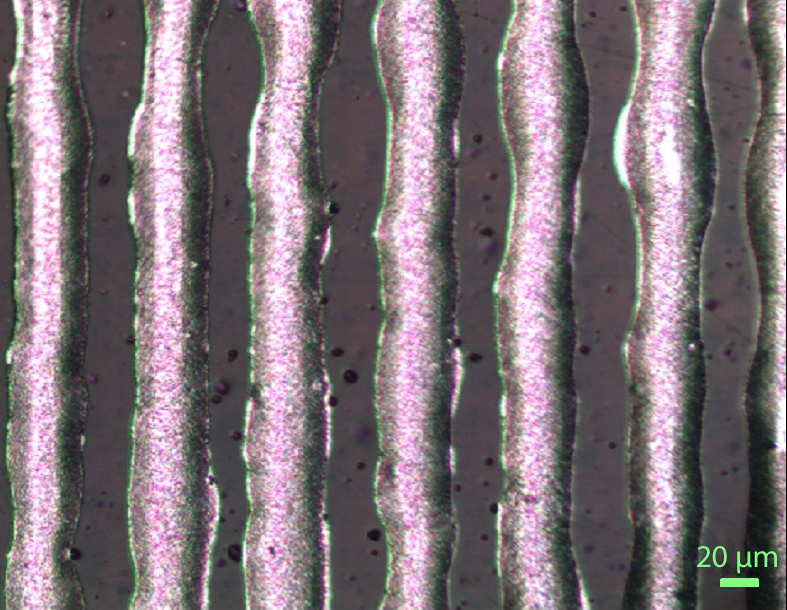}
	  \caption{ \small An image of the 40 $\mu m$ gap, 40 $\mu m$ width polarizer with 4 layers (40G/40W/4L). The period of the polarizer is 80 $\mu m$; however, the gap between the lines is slightly smaller than the expected 40 $\mu m$ due to running of excess silver ink while printing. Implications of this are discussed in the text.}
\end{figure}

\normalsize In TDTS, an infrared femtosecond laser pulse is split into two paths that sequentially excite a pair of photoconductive antenna called "Auston" switches. The first switch generates a mostly vertically-polarized THz pulse, which then travels through the sample. The second antenna receives the THz pulse and measures its electric field \cite{tdts}. 

Calculating the transmission of the polarizer when the printed lines are aligned $T_{pass}$ and perpendicular $T_{polarize}$ to the incoming THz radiation allows for the calculation of the DoP of the polarizer. We define \par

\begin{equation} 
    \label{DoP}
    DoP = \frac{T_{pass} - T_{polarize}}{T_{pass} + T_{polarize}}.
\end{equation}

To perform the measurements, the polarizers were sandwiched between two metal disks each with an 8 mm aperture for their easy alignment. This was placed in a rotation optic stage to allow for manual angle adjustments. When measuring more than one polarizer at a time, multiples of the metal disk and polarizer sandwich were placed in a single rotation optic stage. Conventional wire grid polarizers were placed in front of and behind the polarizer in the THz beam path to remove all horizontal components of the incident electric field. The manual alignment of stacked polarizers was performed under an optical microscope. \par

The TDTS spectrometer on which the measurements were performed reliably measures over the frequency range of 0.2 - 2.2 THz \cite{polarizermathandapp}, however, the polarizers used in this study limit this range. To determine the effective range of measurement, multiple scans for each polarizer geometry were compared. If there was at least 97$\%$ agreement between the scans at a certain frequency, this frequency was considered within the effective range of measurement and the effective range of polarization. This range was determined to be at least 0.3 THz to 1 THz for all polarizers.

\section{RESULTS AND DISCUSSION\vspace{-2ex}}
Measurements were carried out on 8 different polarizers, each with one of 8 possible combinations of the parameters measured. The gap (G) between the printed lines was either 40 $\mu m$ or 80 $\mu m$. The width (W) of the lines was also either 40 $\mu m$ or 80 $\mu m$. The number of printed layers (L) was either 2 or 4. By measuring the various combinations of these polarizers, we determined which parameters were optimal for polarization. \par

As mentioned previously, especially with the 4-layer polarizers, the expected gap size was larger than what was observed. With 80 $\mu m$ width samples there was significant bleeding between ink lines and this likely reduced the transmission and DoP of these polarizers. \par

\begin{figure}[!ht]
	\center
	\includegraphics[width = 0.67\linewidth]{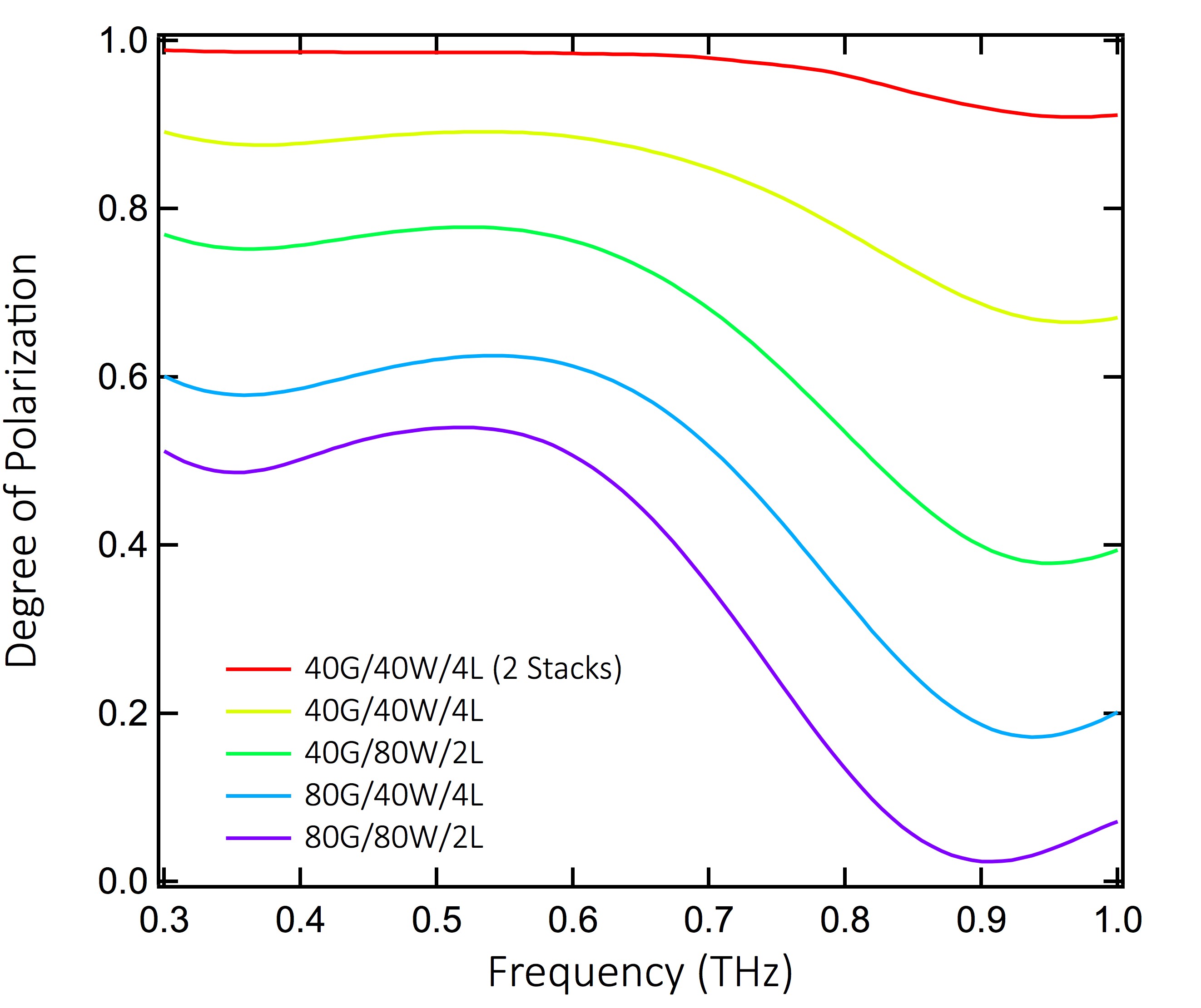}
		\caption{ \small A comparison of the degree of polarization (DoP) of polarizers with different parameters over the effective frequency range. The red line depicts two 40 $\mu m$ gap, 40 $\mu m$ width, 4 layer (40G/40W/4L) polarizers stacked and shows a very high DoP along most of the frequency range.}
\end{figure}

\normalsize Fig. 2 shows a comparison of the DoP as a function of frequency for a sampling of the polarizers measured in this experiment. These polarizers represent the range of quality of polarization and show the effectiveness of stacking polarizers. The red curve depicts two stacked 40 $\mu m$ gap, 40 $\mu m$ width, 4 layer (40G/40W/4L). This DoP is nearly identical to $1 - (1 - DoP)^2$ using the DoP of the 40G/40W/4L polarizer. This indicates that stacked polarizers behave as a series of independent polarizers.  \par

\begin{figure}[!ht]
	\center
	\includegraphics[width = 0.32\linewidth]{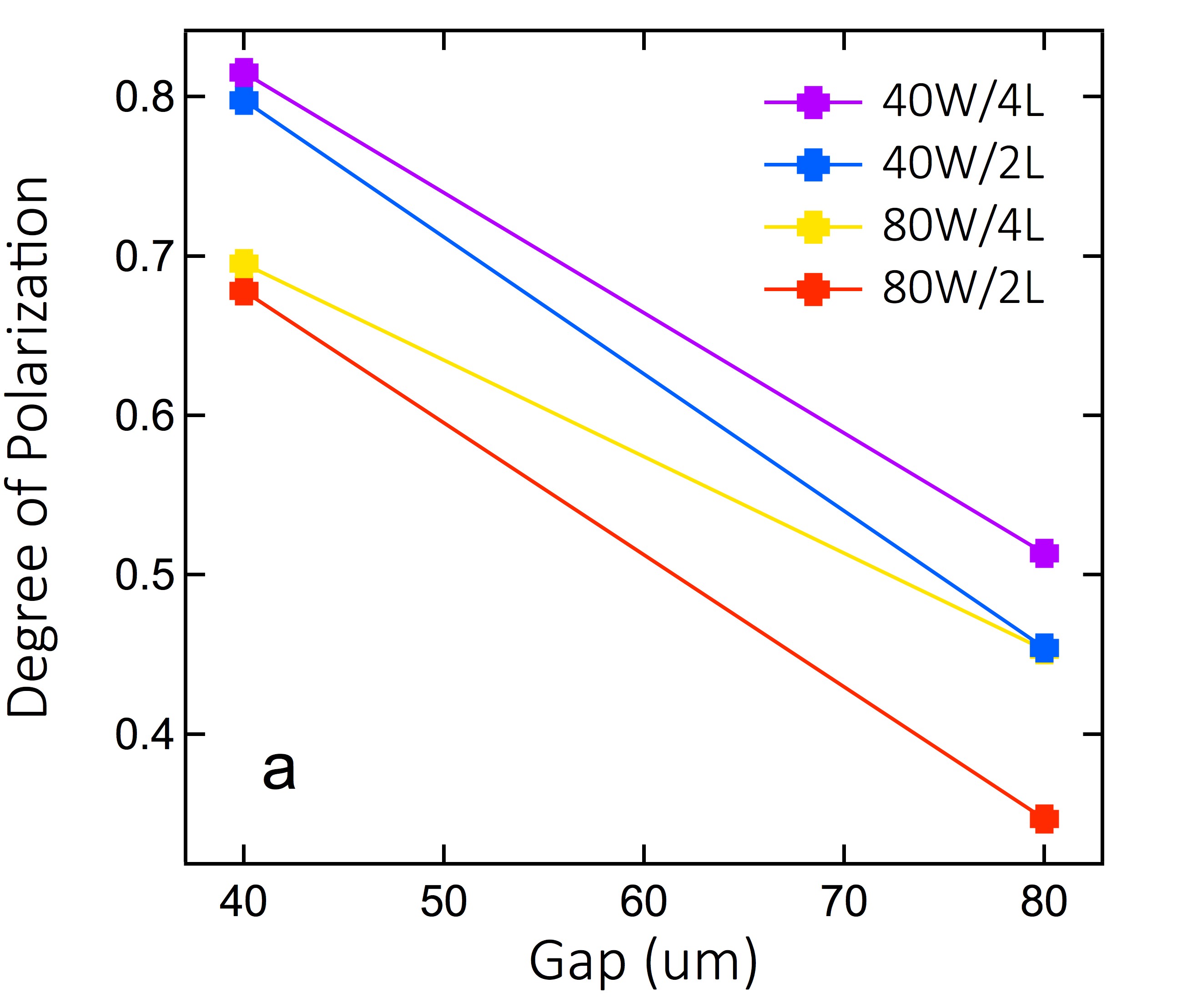}
	\includegraphics[width = 0.315\linewidth]{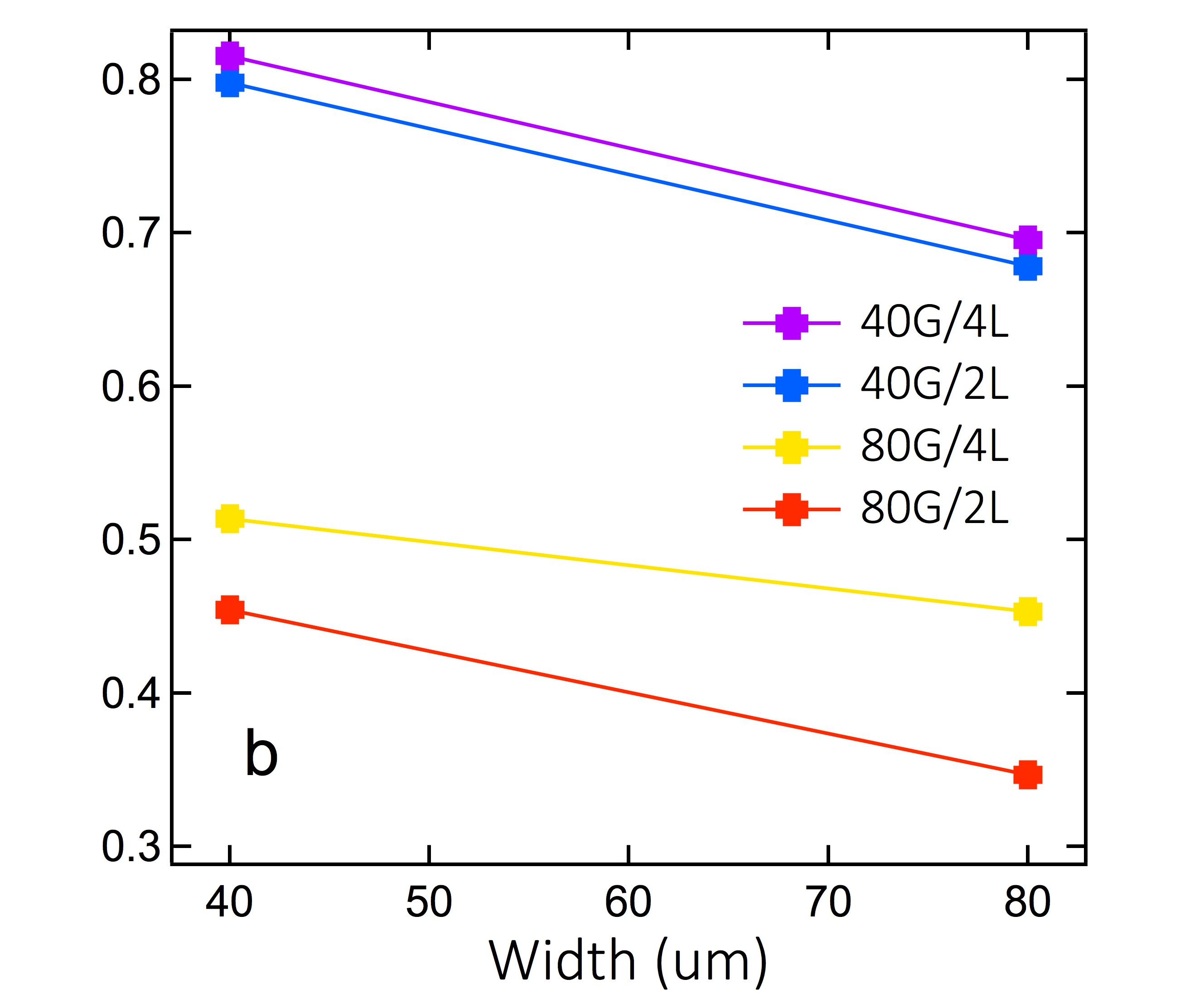}
	\includegraphics[width = 0.32\linewidth]{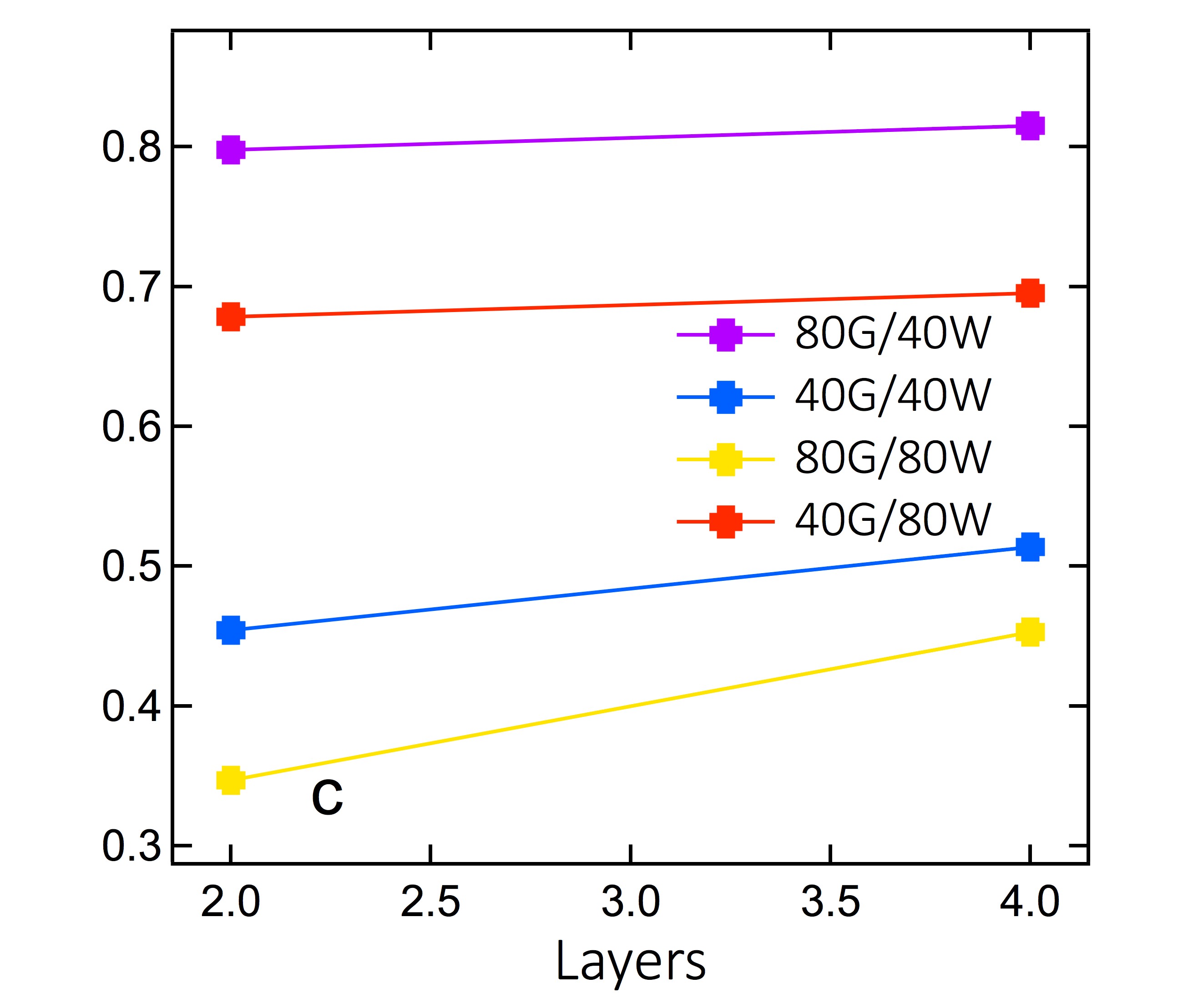}
	  \caption{\small  Degree of polarization (DoP) at 0.7 THz as a function of the (a) gap size, (b) ink line width, and (c) number of printed layers.}
\end{figure}

\normalsize Fig. 3 depicts the result of holding two parameters constant and varying the third. In Fig. 3a the gap between lines is varied and a large DoP dependence is revealed; with a decrease in the gap, there is an increase in DoP. Fig. 3b also shows an inverse relationship between DoP and the width. Fig. 3c shows a slight increase in polarization with an increase in the number of layers printed; however, this may be attributed to the increase in number of layers causing slight ink bleeding and larger ink line width. As long as the ink lines are un-broken and opaque to THz radiation with just 2 layers, the polarization may not be strongly dependent on the number of layers. 
\par
Effective medium theory can be used to model the dependence of properties of the polarizers. Here we use the same approach as detailed in Ref. \cite{analysis}.  First, we define $T_{polarize}$ and $T_{pass}$: $T_{polarize} \approx \frac{4n_{\parallel}}{(1+n_{\parallel})^2}$,
$T_{pass} \approx \frac{4n_{\bot}}{(1+n_{\bot})^2}$. Here, $n_{\parallel}$ is the effective index of refraction when the polarizer's conductive ink lines are parallel to the eletric field of the THz radiation, and $n_{\bot}$ is the effective polarizer with perpendicular allignment. By approximating the ink lines to be long rectangular prisms and the wavelength of the THz radiation to be much greater than the period ($G + W$) of the polarizers, we can use a low order approximation given by the two following equations,

\begin{equation}
    \label{parallel index of refraction}
	n_{\parallel} = \sqrt{n_{0}^2(1-D) + n_m^2D}
\end{equation}
\begin{equation}
    \label{perp index of refraction}
    n_{\bot} = \sqrt{\frac{n_0^2 n_m^2}{n_m^2(1-D) + n_0^2D}}
\end{equation}

where $D = \frac{W}{G + W}$, $n_m$ is the index of refraction of the ink, and $n_0$ is the index of refraction of the material that is between the ink lines.  Using these equations, it can be shown that $n_{\parallel} \approx n_{m} \sqrt{D}$ and $n_{\bot} \approx \frac{n_0}{\sqrt{1-D}}$, $n_0 \approx 1$. We plug back into Eq. (1) and find that

\begin{equation}
    \label{perp index of refraction}
    DoP \approx \frac{\sqrt{G}(G+W) + n_m^2 W \sqrt{G} - n_m G \sqrt{W} - n_m\sqrt{W}(G+W)}{\sqrt{G}(G+W) + n_m^2 W \sqrt{G} + n_m G \sqrt{W} + n_m\sqrt{W}(G+W) + 4n_m\sqrt{GW(G+W)}}
\end{equation}

By fixing $W$ and increasing $G$, we find that DoP decreases, and by fixing $G$ and increasing $W$, DoP decreases as well for reasonable values of $n_m$. Additionally, this equation implies that the DoP should decrease more quickly for increasing $G$ than for increasing $W$. This is consistent with our results.

\begin{figure}[!ht] 
	\center
	\includegraphics[width=0.49\linewidth]{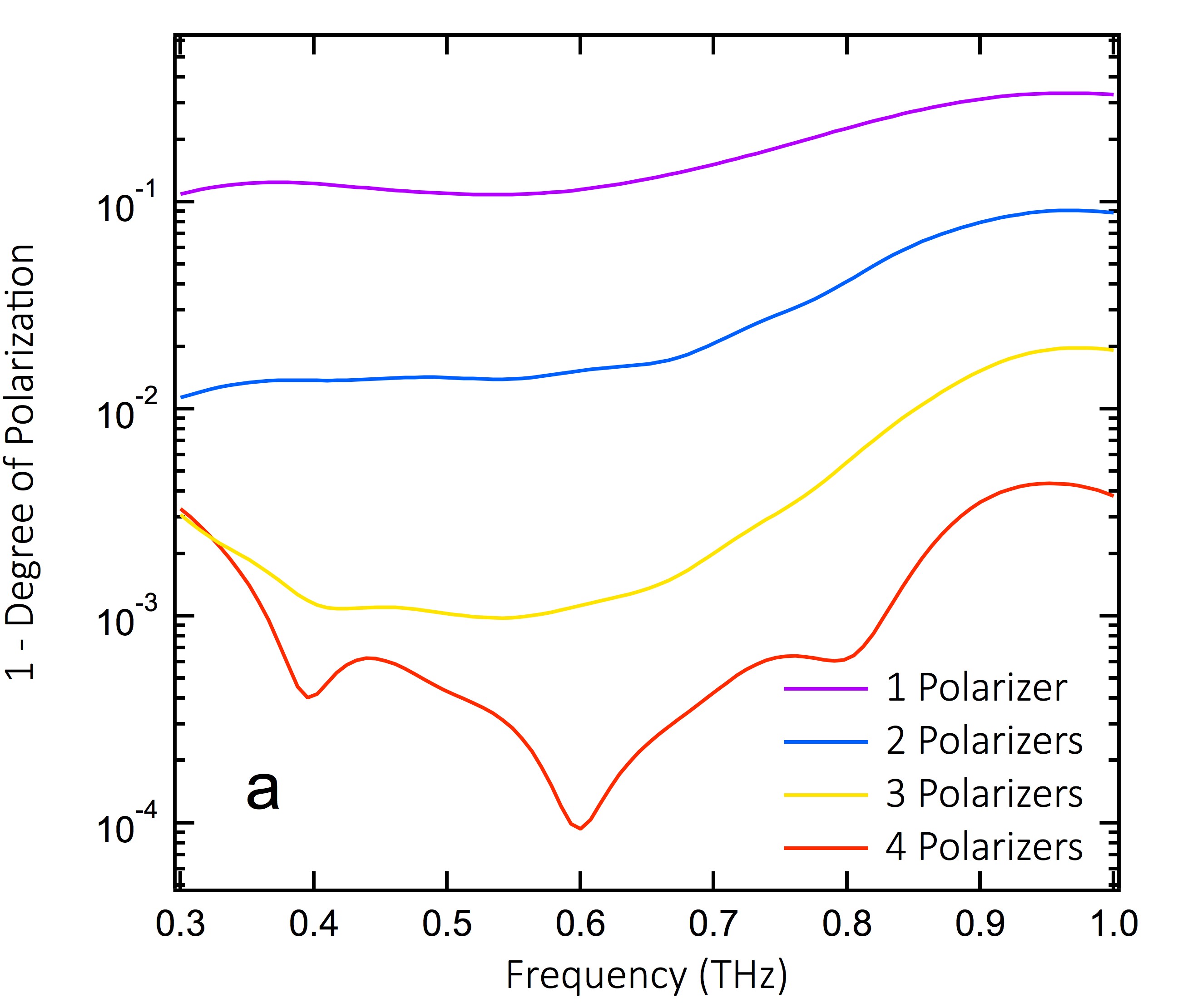}
	\includegraphics[width=0.49\linewidth]{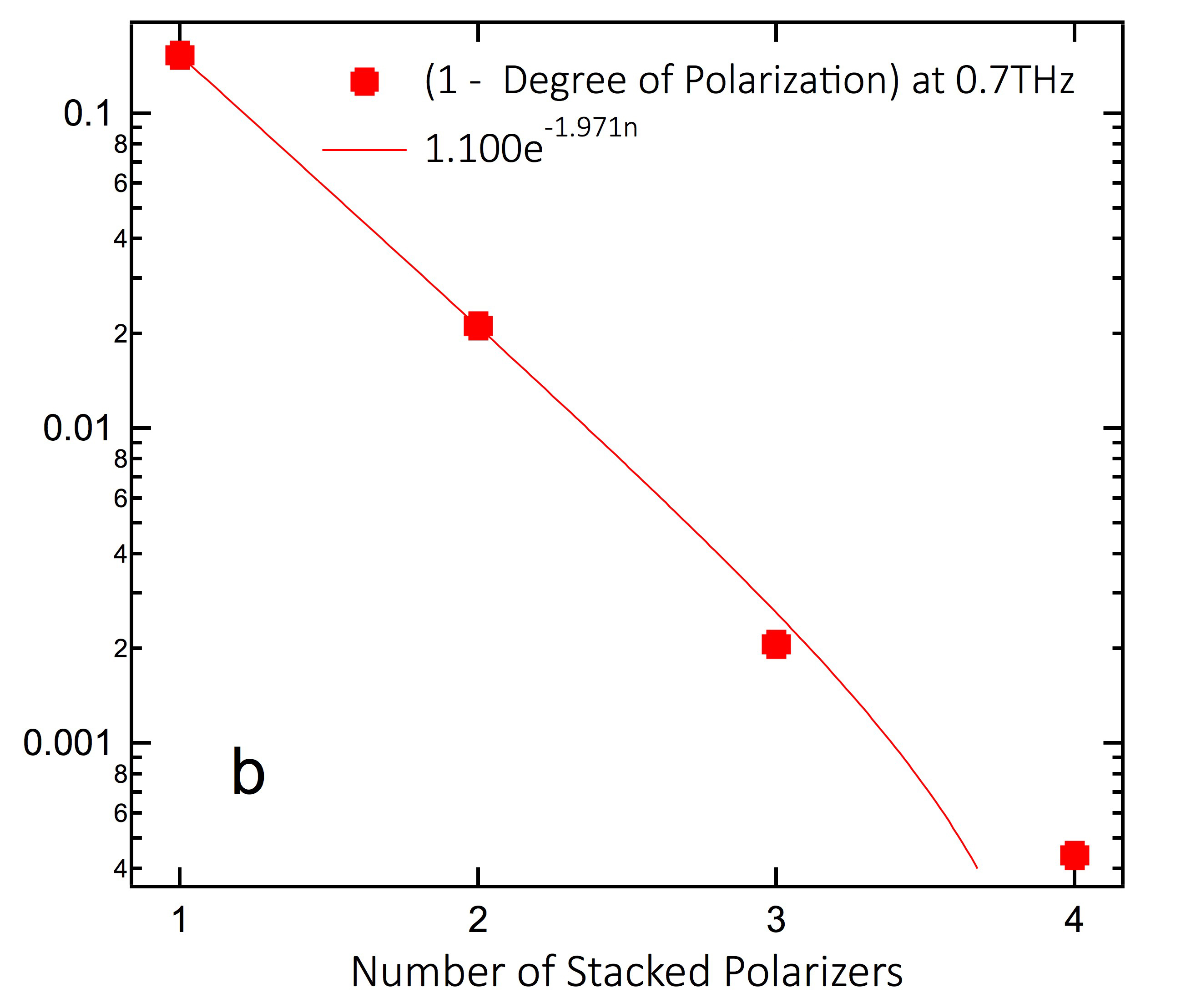}
	  \caption{\small (a) Plot showing the effect on 1-DoP of stacking 40G/40W/4L polarizers (the DoP increases as the number of polarizers increases). (b) 1-DoP at 0.7 THz as a function of the number of stacked polarizers. Each additional polarizer reduces the DoP by nearly one order of magnitude.}
\end{figure}

\normalsize Fig. 4 depicts the result of investigating the effect on DoP from stacking the best individual polarizer: 40G/40W/4L. Fig 4a shows the result of stacking polarizers on the effective THz frequency range. For each polarizer added, 1-DoP is reduced by nearly one order of magnitude across the entire effective frequency range. This proceeds until the amount of light passed through starts approaching the noise level, which occurred with 4 stacked polarizers. A frequency cut at 0.7 THz was taken and plotted against the DoP (Fig 4b). There is a linear relationship between the log of the DoP and the number of stacked polarizers. Assuming this linear relationship holds up, we can predict with reasonable accuracy the DoP for larger numbers of polarizers. Additionally, fitting implies a 86$\%$ decrease in 1-DoP for each additional polarizer. The inaccuracy in the fit for $n = 4$ could be due to the measurement of 4 stacked polarizers nearing the noise floor of our measurement technique. \par

Since the polarizers were printed on Kapton films, and this material is nearly transparent in the THz frequency range, there is little reduction in transmission due to the polarizers. At 0.7 THz, the highest and lowest measured transmissions were the 40W and 4L polarizers with transmissions of 0.88 $\pm$ 0.02 and 0.74 $\pm$ 0.17, respectively. The transmission for a single 40G/40W/4L polarizer at 0.7 THz is 0.89 $\pm$ 0.01, with an average transmission across the effective frequency range of 0.93 $\pm$ 0.02. Using this value to calculate the expected transmission for 4 stacks of the 40G/40W/4L polarizers results in a transmission of 0.75. This is similar to the measured value of 0.792 $\pm$ 0.003. Since these polarizers are inexpensive and easy to make with the correct machinery, 10-15 of these polarizers can be stacked to achieve ultra high extinction ratios. Stacking, for example, 10 of these polarizers may only reduce transmission by 56$\%$ and could have an extinction ratio of over $2 \times 10^{8}$:1 instead of the 2300:1 achieved with 4 polarizers. This far surpasses any individual THz range polarizer and could, with correct alignment, be set up in a single optic much like a wire-grid polarizer. This stack of polarizers would be less fragile, have ultra high DoP/extinction ratio, and cost much less than current commercial polarizers.

\section{CONCLUSION\vspace{-2ex}}

We have characterized the polarizing capability in the THz frequency range of novel polarizers made from silver-nanoparticle ink printed on Kapton film. We found that a decrease in the gap between ink lines, decrease in the ink line width, and increase in the number of layers results in a higher degree of polarization within the parameter limits of our experiment and method used for printing. We demonstrated that these simple polarizers when stacked are comparable to commercial polarizers. The polarizer topology made of lines of a gap of 40 $\mu m$ and a width of 40 $\mu m$ is highly effective in the 0.3-1 THz frequency range. Further reduction in gap size may also increase the degree of polarization, and further stacking of these polarizers may demonstrate an ultimate capability for polarization.\par

The THz instrumentation development was funded by the Gordon and Betty Moore Foundation through Grant GBMF2628 to NPA. The authors would like to thank the National Science Foundation, the Defense Threat Reduction Agency, and the Semiconductor Research Corporation for their support with the printing method used in this work.

\end{document}